\documentclass[twocolumn,preprintnumbers,amssymb,amsmath,superscriptaddress,letterpaper]{revtex4}[10pt]

\usepackage{graphicx}
\usepackage{dcolumn}
\usepackage{bm}
\usepackage{pstricks}
\usepackage{epsfig}

\newcommand{\be}{\begin{equation}}
\newcommand{\ee}{\end{equation}}
\newcommand{\bea}{\begin{eqnarray}}
\newcommand{\eea}{\end{eqnarray}}

\begin{document}
%\draft

\title{New Constraints from PAMELA anti-proton data on Annihilating and Decaying Dark Matter}

\author{Ilias Cholis}
\email{ilias.cholis@sissa.it}
\affiliation{Astrophysics Sector, La Scuola Internazionale Superiore di Study Avanzati and Instituto Nazionale di Fisica Nucleare, Sezione di Trieste, via Bonomea 265, 34136 Trieste Italy}
\affiliation{Center for Cosmology and Particle Physics, Department of Physics, New York University, 
New York, NY 10003}

\date{\today}

\begin{abstract}
Recently the \textit{PAMELA} experiment has released its updated anti-proton flux and anti-proton to proton flux ratio data up to energies of $\approx$ 200GeV.  With no clear excess of cosmic ray anti-protons at high energies, one can extend constraints on the production of anti-protons from dark matter. In this letter, we consider both the cases of dark matter annihilating and decaying into standard model particles that produce significant numbers of anti-protons. We provide two sets of constraints on the annihilation cross-sections/decay lifetimes. In the one set of constraints we ignore any source of anti-protons other than dark matter, which give the highest allowed cross-sections/inverse lifetimes. In the other set we include also anti-protons produced in collisions of cosmic rays with interstellar medium nuclei, getting tighter but more realistic constraints on the annihilation cross-sections/decay lifetimes.
\end{abstract}

\maketitle

\section{Introduction}
\label{sec:intro}

Dark matter (DM) composes approximately $85 \%$ of the matter density of the universe, but its particle physics properties in the WIMP case still remain unknown. Recently new measurements of cosmic rays\cite{Boezio:2008mp, Adriani:2008zr, Adriani:2008zq,Collaboration:2008aaa, Aharonian:2009ah, Abdo:2009zk, ATIClatest} have generated new model building and new constraints on dark matter properties. At the same period direct detection experiments \cite{Baudis:2007ew, Aprile:2009yh, Aalseth:2010vx, Ahmed:2009zw, Bernabei:2008yh} have provided their own set of constraints on the mass and interaction properties of DM particles with nucleons \cite{Angle:2008we, Kopp:2009qt, Fitzpatrick:2010em, Chang:2010yk, Bernabei:2010mq, Aprile:2010um, Collar:2010gg}.

The decay or annihilation products of dark matter generally yield equal amounts of matter and antimatter.\footnote{For models that annihilate asymmetrically to antimatter vs matter see \cite{Feldstein:2010xe, LisaSpencer}.} In contrast, antimatter cosmic rays (CR) that are produced by Standard model interactions of high energy protons and nuclei with the Interstellar Medium (ISM) have fluxes that are significantly lower than their matter CR antiparticles at a broad range of energies. Thus antimatter to matter ratios such as the positron fraction $e^{+}/(e^{+} + e^{-})$ or the $\bar{p}/p$ ratio can provide a very valuable probe for identifying DM or constraining its properties. 

\cite{Adriani:2010rc} has provided an update to \cite{Adriani:2008zq} of the anti-proton to proton flux ratio measurement by the \textit{PAMELA} \cite{Boezio:2008mp} satellite, including higher statistics and a new higher energy data point extending the measurement of the $\bar{p} / p$ flux ratio up to $\approx 200$ GeV. This high energy region has not been probed before with such accuracy,\footnote{ARGO-YBJ experiment has put upper limits to the $\bar{p} / p$ flux ratio at $\sim$ TeV region \cite{DiSciascio:2009ir} but their allowed  $\bar{p} / p$ ratio values are very high ($O(0.1)$).} and thus provides us with the opportunity of constraining the production rate of $\bar{p}$ from DM and to probe new parts of the parameter space for the hadronic component of the annihilation/decay products of DM. 
In \cite{Donato:2008jk, Cirelli:2008pk} the authors have provided constraints on annihilating DM via modes that produce significant fluxes of $\bar{p}$s using the \textit{PAMELA} $\bar{p}$ up to 100 GeV\cite{Adriani:2008zq}.

In this letter we present the \textit{new} limits on annihilating DM cross-sections or decaying DM lifetimes, into $W^{+}W^{-}$, $ZZ$ and $b \bar{b}$ for a region of DM masses $m_{\chi}$. 

%%%%%%%%%%%%%%%%%%%%%%%%%%%%%%%%%%%%%%%
%%%%%%%%%%%%%%%%%%%%%%%%%%%%%%%%%%%%%%%
\section{Assumptions}
\label{sec:assumptions}

We consider as reference annihilation mode: $\chi \chi \rightarrow  W^{+}W^{-}$ and as reference decay mode: $\chi \rightarrow  W^{+}W^{-}$, and study the region 100 GeV $< m_{\chi} < 10$ TeV for annihilation and 200 GeV $< m_{\chi} < 20$ TeV for decay. We use PYTHIA \cite{Sjostrand:2006za} to get the $\bar{p}$ injection spectra into the ISM from the decay and hadronization of the boosted W's. 

High energy electrons and protons loose energy via inverse Compton scattering on CMB, IR and starlight photons, synchrotron radiation for which the energy loss rates $\frac{dE}{dt}$ scale as $\gamma^{2}$,\footnote{Assuming validity of the Thomson cross-section.} ($\gamma$ being the Lorentz boost) and bremsstrahlung for which $\frac{dE}{dt} \sim \gamma$. Thus high energy $p$ and $\bar{p}$ have much lower energy losses compared to equal energy $e^{\pm}$. On the other hand, high energy protons and electrons of energy $E \gg m_{p}c^{2}$ have identical rigidities and thus diffuse in the ISM in the same way. As a result, high energy protons observed locally originate from a bigger volume than electrons of the same energy in general. Thus the DM profile distribution is of importance. We show the induced constraints from the new $\bar{p}/p$ flux ratio and $\bar{p}$ flux data for a cuspy profile which we take to be Einasto\cite{Einasto},
\begin{equation}
\label{eq:Einasto}
	\rho(r)=\rho_{0} \exp\left[-\frac{2}{\alpha}\left(\frac{r^{\alpha}-R_{\odot}^{\alpha}}{r_{-2}^{\alpha}}\right)\right],
\end{equation}
and a cored profile which we take to be a cored isothermal profile described by:
\begin{equation}
\label{eq:CoredIsothermal}
	\rho(r)=\rho_{0}\frac{R_{\odot}^{2}+R_{c}^{2}}{r^{2}+R_{c}^{2}}.
\end{equation}
The parameters values that we assume are $\alpha = 0.17$ and $r_{-2}=25 \; \rm kpc$, following Merritt et al. \cite{Merritt:2005xc} and $R_{c} = 2.8\; \rm kpc$ \cite{Moskalenko:1999sb, Kamionkowski:1997xg}.
We take the value of the local density to be $\rho_{0}=0.4 \; \rm GeV cm^{-3}$~\cite{Catena:2009mf} and $R_{\odot} = 8.5\; \rm kpc$.

For the propagation of CRs in the Galaxy we use the publicly available GALPROP\cite{Strong:2007nh} with  diffusion within a zone of half width of 4 kpc. Diffusion, energy loss, and diffusion zone parameters are such that predicted spectra of the high energy nuclei fluxes are consistent with local CR measurements.

To determine upper limits on DM annihilation cross-sections and decaying DM inverse lifetimes, we first ignore other astrophysical contributions to the $\bar{p}$ flux, and thus extract upper limits by comparing the \textit{PAMELA} $\bar{p}$ flux to the $\bar{p}$ flux from DM.

Yet, we know anti-protons are produced also by high energy protons colliding to ISM protons or nuclei, thus it would be an oversimplification to ignore any background. Furthermore,  anti-protons originating from DM give a significantly harder spectrum than that observed at $E_{\bar{p}} > 10$ GeV (with the exception of $m_{\chi} < 500$ GeV(1 TeV) for annihilation(decay)), failing to fit the observed spectrum if they are the dominant component at $E_{\bar{p}} < 100$ GeV. Thus we extract a second set of constraints assuming the high energy CRs contribution to the  $\bar{p}$ flux. Using the same propagation model as the one we assumed for the DM originated $\bar{p}$ fluxes we modified the GALPROP assumptions for the $pp$ cross-sections into $\bar{p}$ \cite{Moskalenko:2001ya}, that describes the secondary anti-protons component which is the dominant component on the flux at high $\bar{p}$ energies. In our assumptions like the authors of \cite{Moskalenko:2001ya} we fit the ratio of $\bar{p}$ yield from $pp$ reactions to the Simon et al. data \cite{1998ApJ...499..250S} in order to have agreement up to the highest $T_{\bar{p}} \approx 100$ GeV available (where $T_{\bar{p}}$ is the $\bar{p}$ kinetic energy). Our fitted ratio is given by:
\begin{equation}
\label{eq:OurppRatio}
	\frac{\sigma_{\Sigma}}{\sigma_{pp}} = 0.12 \left(\frac{T_{\bar{p}}}{GeV}\right)^{-1.6} - 1.5\times 10^{-3}\frac{T_{\bar{p}}}{GeV} + 1.8.
\end{equation} 
GALPROP assumes for the same ratio 
\cite{Moskalenko:2001ya}:\footnote{http://galprop.stanford.edu/web$\_$galprop/galprop$\_$home.html .}

\begin{equation}
\label{eq:GalpropppRatio}
	\frac{\sigma_{\Sigma}}{\sigma_{pp}} = 0.12 \left(\frac{T_{\bar{p}}}{GeV}\right)^{-1.67} + 1.78.
\end{equation} 
Our assumptions provide a lower ratio of $\sigma_{\Sigma}/\sigma_{pp}$ than \cite{Moskalenko:2001ya} at $T_{\bar{p}} > 30$ GeV and slightly higher ratio at $5-10$ GeV getting a slightly better agreement to the Simon et al. high energy data. 

As lower dark matter masses can contribute in the $\bar{p}$ fluxes down to energies of $\sim GeV$ we allow for the background $\bar{p}$ flux normalization to be decreased up to $10 \%$ from the best fit value in the absence of a DM originating component. 

%%%%%%%%%%%%%%%%%%%%%%%%%%%%%%%%%%%%%%%
%%%%%%%%%%%%%%%%%%%%%%%%%%%%%%%%%%%%%%%
\section{Results}
\label{sec:results}

\begin{figure*}[t]
\centering\leavevmode
\includegraphics[width=3.20in,angle=0]{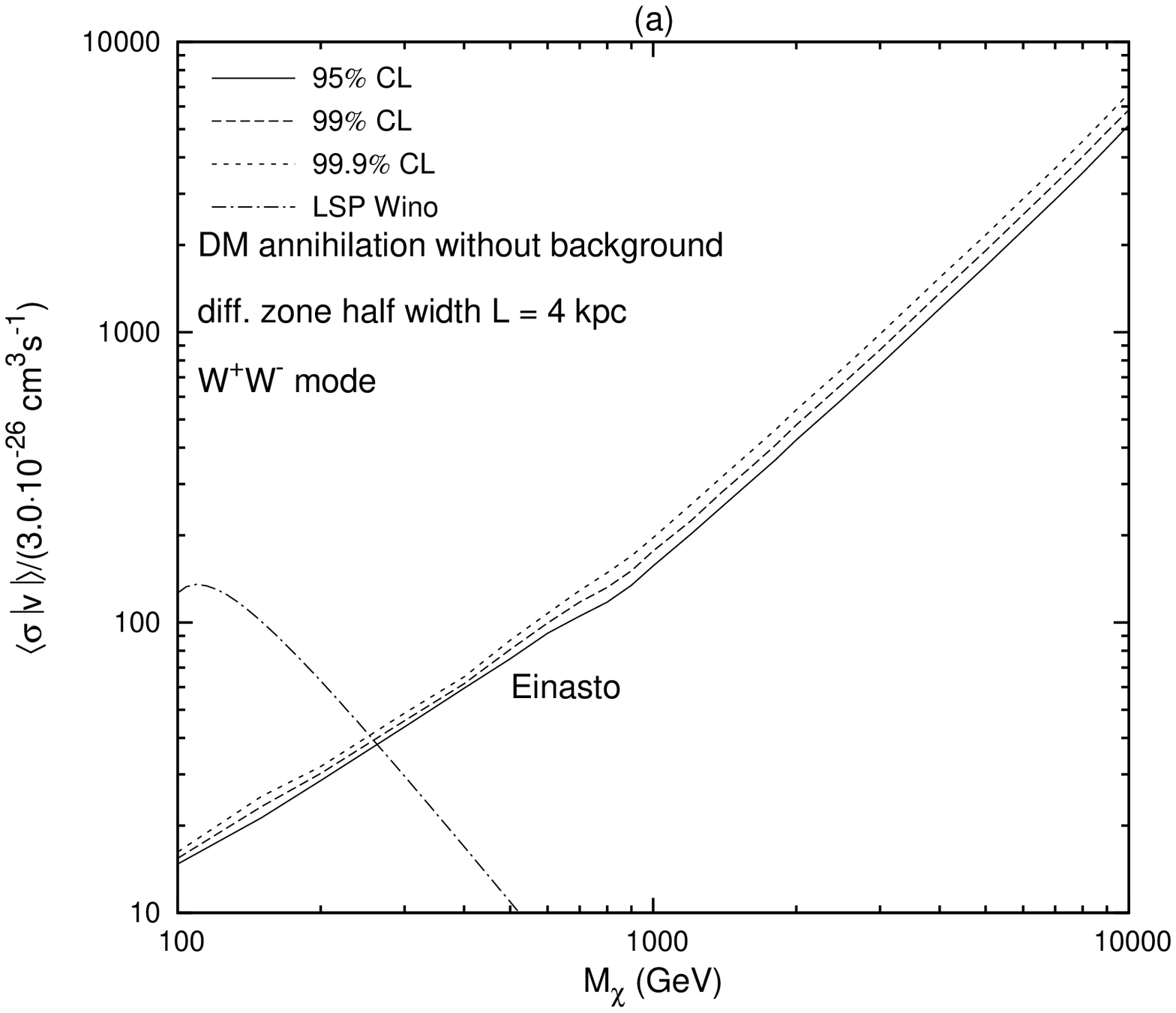}
\hspace{-0.1cm}
\includegraphics[width=3.20in,angle=0]{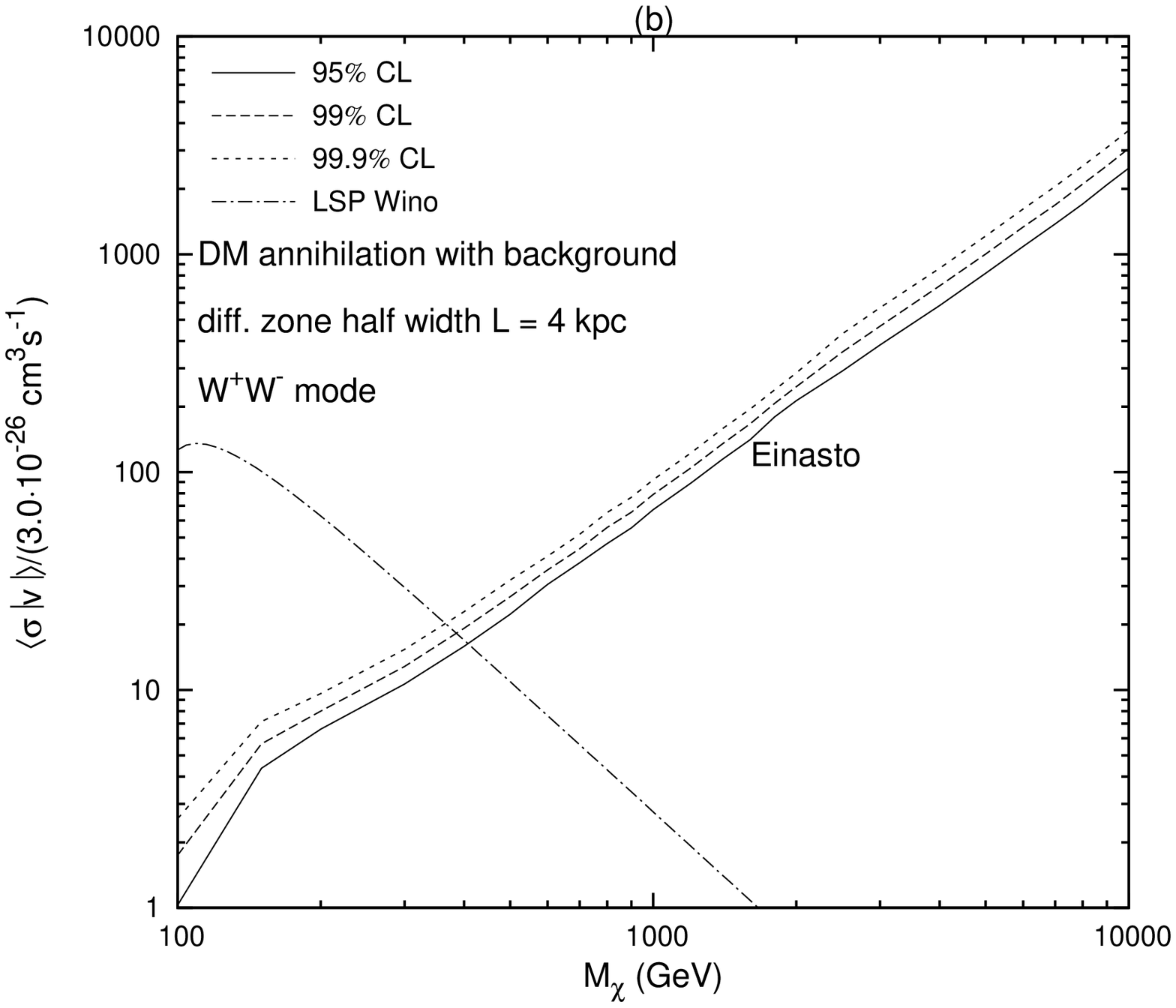}\\
\includegraphics[width=3.20in,angle=0]{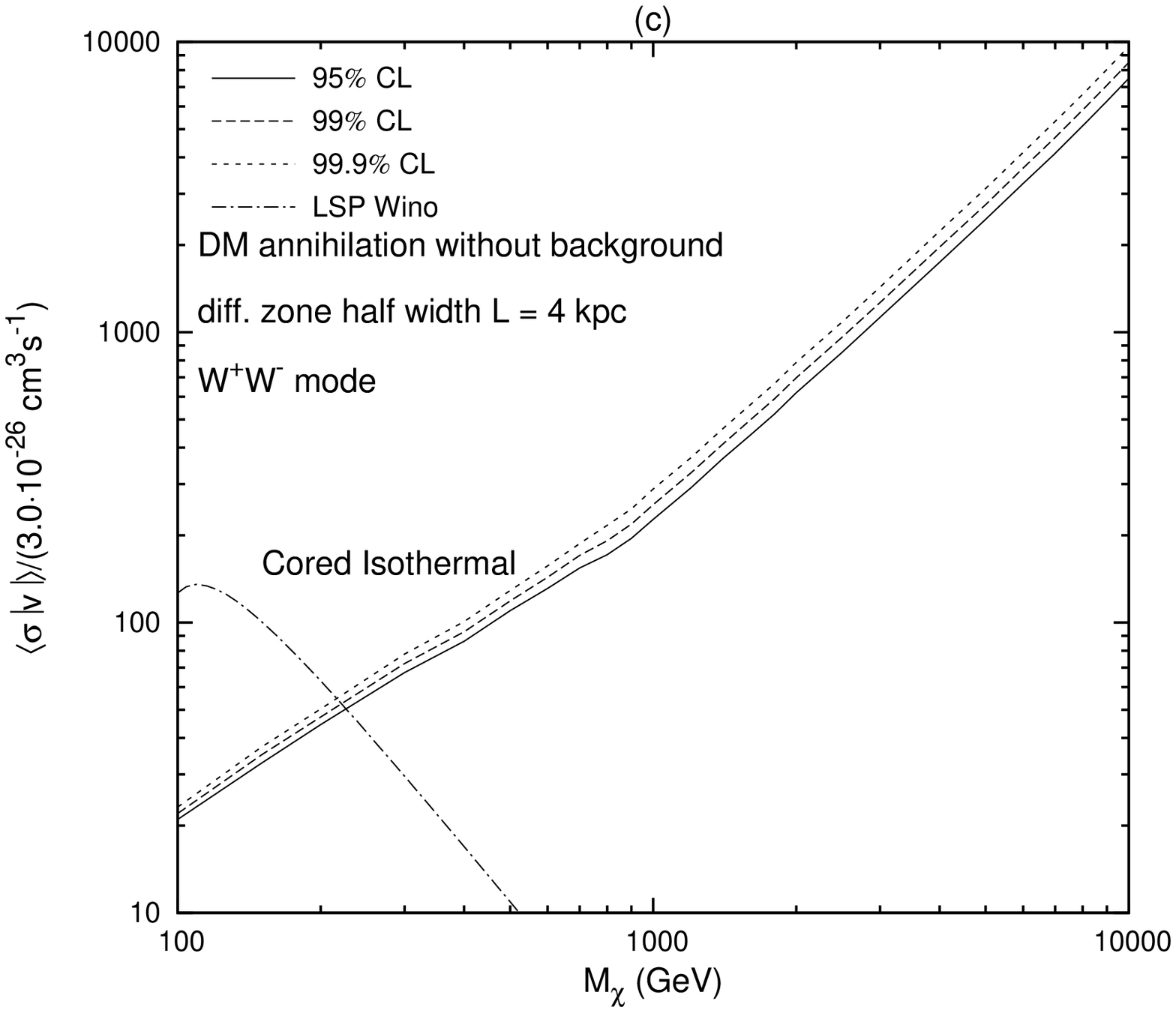}
\hspace{-0.1cm}
\includegraphics[width= 3.20in,angle=0]{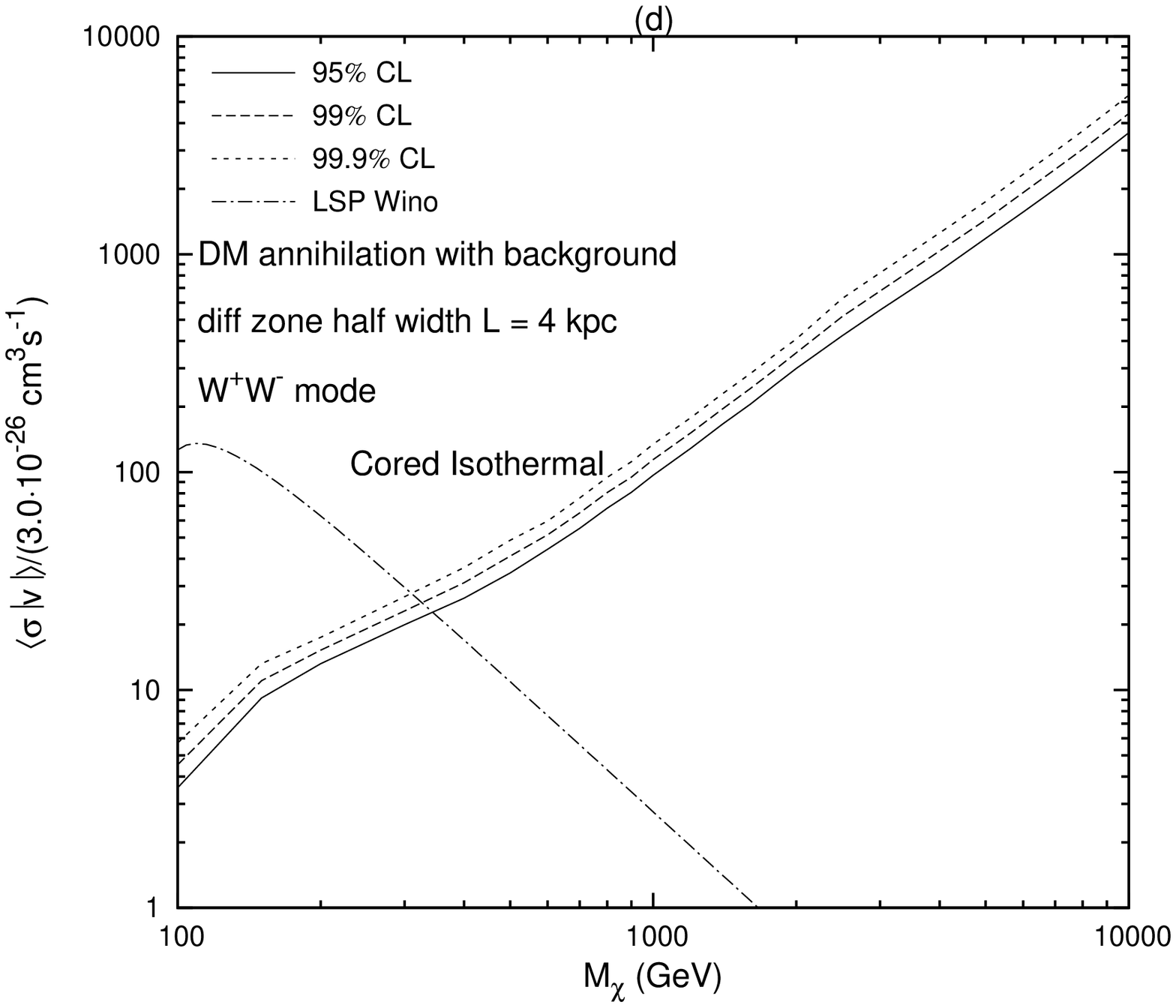}
\caption{Upper limits on annihilation cross-section ratios $\langle \sigma \mid v \mid \rangle$/($3.0\times 10^{-26} cm^{3} s^{-1}$) vs $m_{\chi}$ for $\chi \chi \rightarrow W^{+} W^{-}$, for Einasto (upper row) and cored isothermal profiles (lower row). Left column: ignoring $\bar{p}$ background contribution, right column: including background $\bar{p}$s. Solid lines: $95 \%$, dashed lines: $99 \%$, dotted lines: $99.9 \%$ C.L.s. The LSP Wino s-wave annihilation cross-section is also given \cite{Acharya:2008bk} for reference.}
\label{annihplot}
\end{figure*}

In figure~\ref{annihplot} we present for the case of annihilating DM $\chi \chi \rightarrow W^{+}W^{-}$ the allowed within $95 \%$,  $99 \%$ and $99.9 \%$ C.L.  annihilation   cross-section  ratios  (boost  factors)  $\langle \sigma \mid v \mid \rangle$/($3.0\times 10^{-26} cm^{3} s^{-1}$). We show the results for Einasto profile figure~\ref{annihplot}a,b and for cored isothermal~\ref{annihplot}c,d. Both limits that come in the absence of a background~\ref{annihplot}a,c and in the presence of a background~\ref{annihplot}b,d are shown as described in~\ref{sec:assumptions}. 
We also give as reference the LSP Wino s-wave annihilation cross-section\cite{Acharya:2008bk}.
For $m_{\chi} > 800$ GeV the constraints are imposed from the highest energy data points, while for lower masses the data points in the region of $20-100$ GeV impose the constraints on the annihilation cross-section, as the details of the $\bar{p}$ spectrum at those energies start having an effect on the fit.

Since the background that we use fits well the $\bar{p} / p$ data points up to 30 GeV, while gives lower $\bar{p}$ fluxes at higher energies than the \textit{PAMELA} data suggest, the induced constraints on $\langle \sigma \mid v \mid \rangle$ are more sensitive on background assumptions for the lower masses $m_{\chi}$. For masses $m_{\chi} > 1$ TeV, the presence of a background results in the allowed cross-sections being smaller by a factor of 2, while for lower masses this factor increases up to $\sim$10. 

The difference in assuming an Einasto versus a cored isothermal profile on the allowed cross-sections is a factor of less than 2 at $m_{\chi} > 1$ TeV, and up to a factor of $\sim$3 at the lowest masses, with the Einasto profile setting tighter constraints due to its cuspy DM distribution. 

In table~\ref{annih-table} we present the allowed within $95 \%$ C.L. annihilation cross-section ratios $\langle \sigma \mid v \mid \rangle$/($3.0\times 10^{-26} cm^{3} s^{-1}$) for three different masses, $m_{\chi} = $ 100 GeV, 1.0 TeV and 10 TeV for both Einasto and cored isothermal profiles, and three different annihilation modes: $\chi \chi \rightarrow Z Z$, $\chi \chi \rightarrow W^{+} W^{-}$, $\chi \chi \rightarrow b \bar{b}$. We give the upper limits calculated both including our background and without it (in parentheses). We note also that the limits from the new data are -when using only \textit{PAMELA} data to avoid systematic uncertainties between different experiments-, a factor of 4 (for $\chi \chi \rightarrow W^{+} W^{-}$) to 7 (for $\chi \chi \rightarrow Z Z$) lower for $m_{\chi} = 100$ GeV; and a factor of $\sim 2$ lower for the higher masses\footnote{With the $\bar{p}/p$ data from \cite{Adriani:2008zq} our limit on the BF for a $\chi \chi \rightarrow W^{+} W^{-}$ ($ZZ$) was 95(200) at 1 TeV and 7600(10300) at 10 TeV.}. 

\begin{figure*}[t]
\centering\leavevmode
\includegraphics[width=3.20in,angle=0]{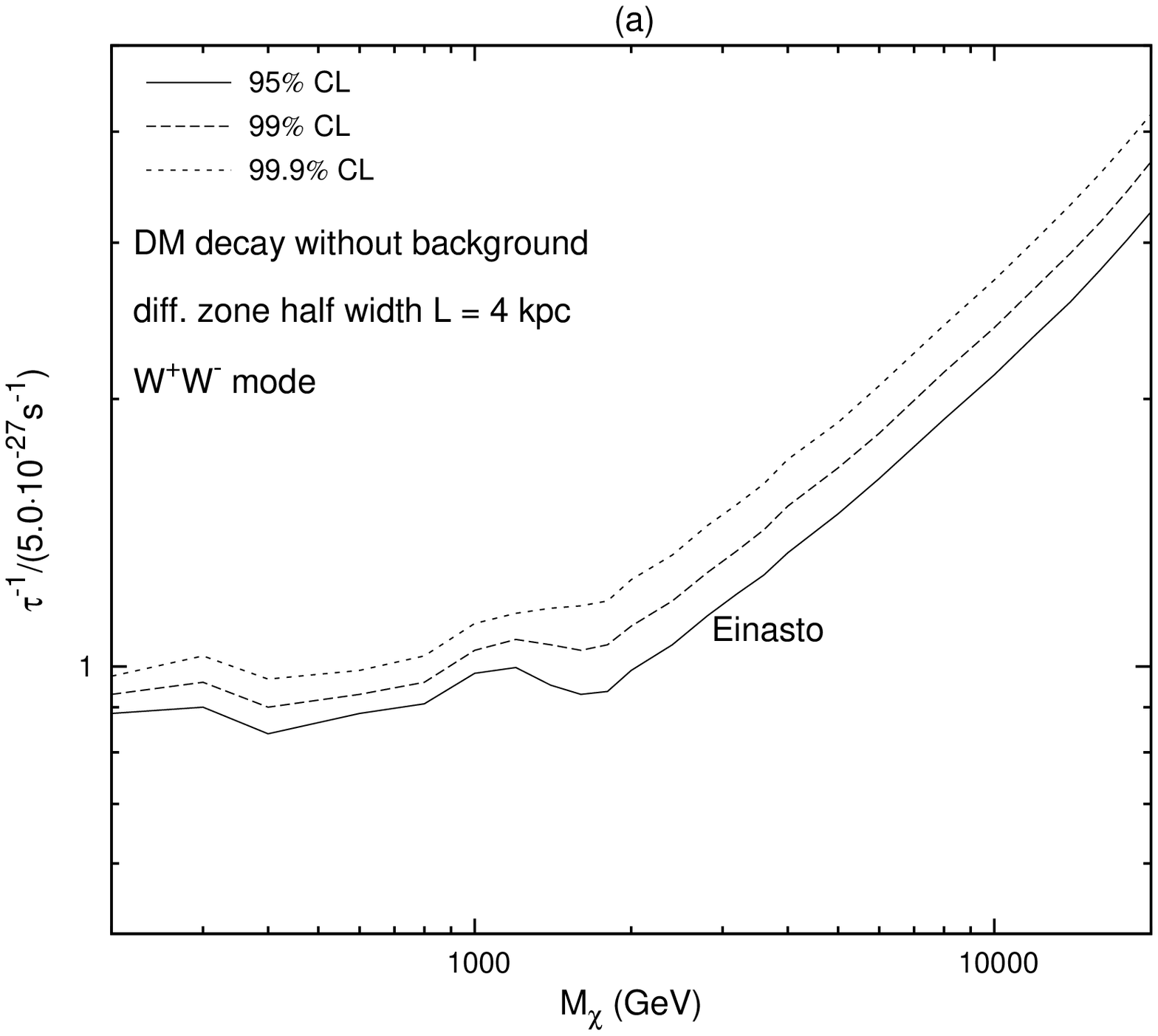}
\hspace{-0.1cm}
\includegraphics[width=3.20in,angle=0]{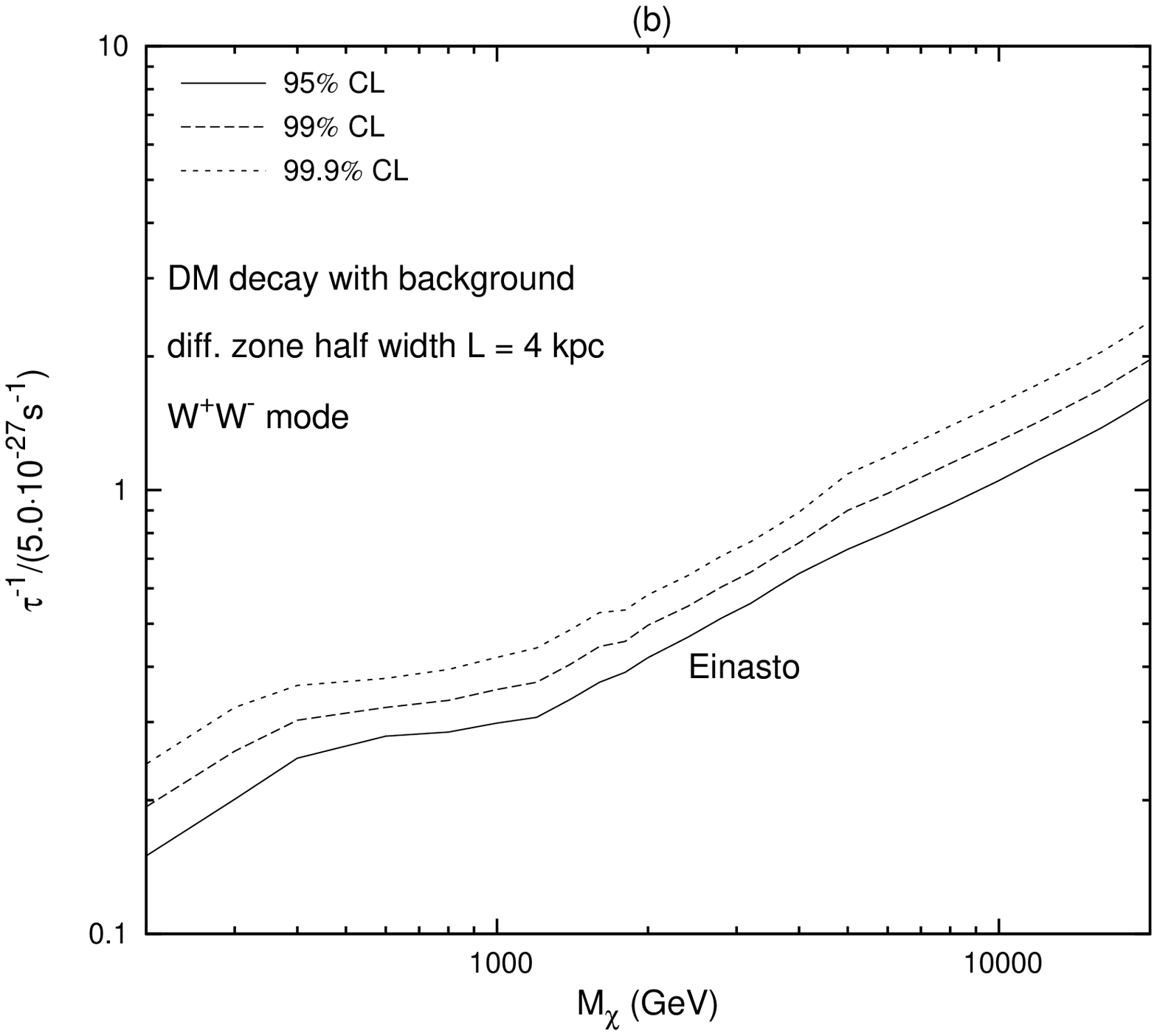}\\
\includegraphics[width=3.20in,angle=0]{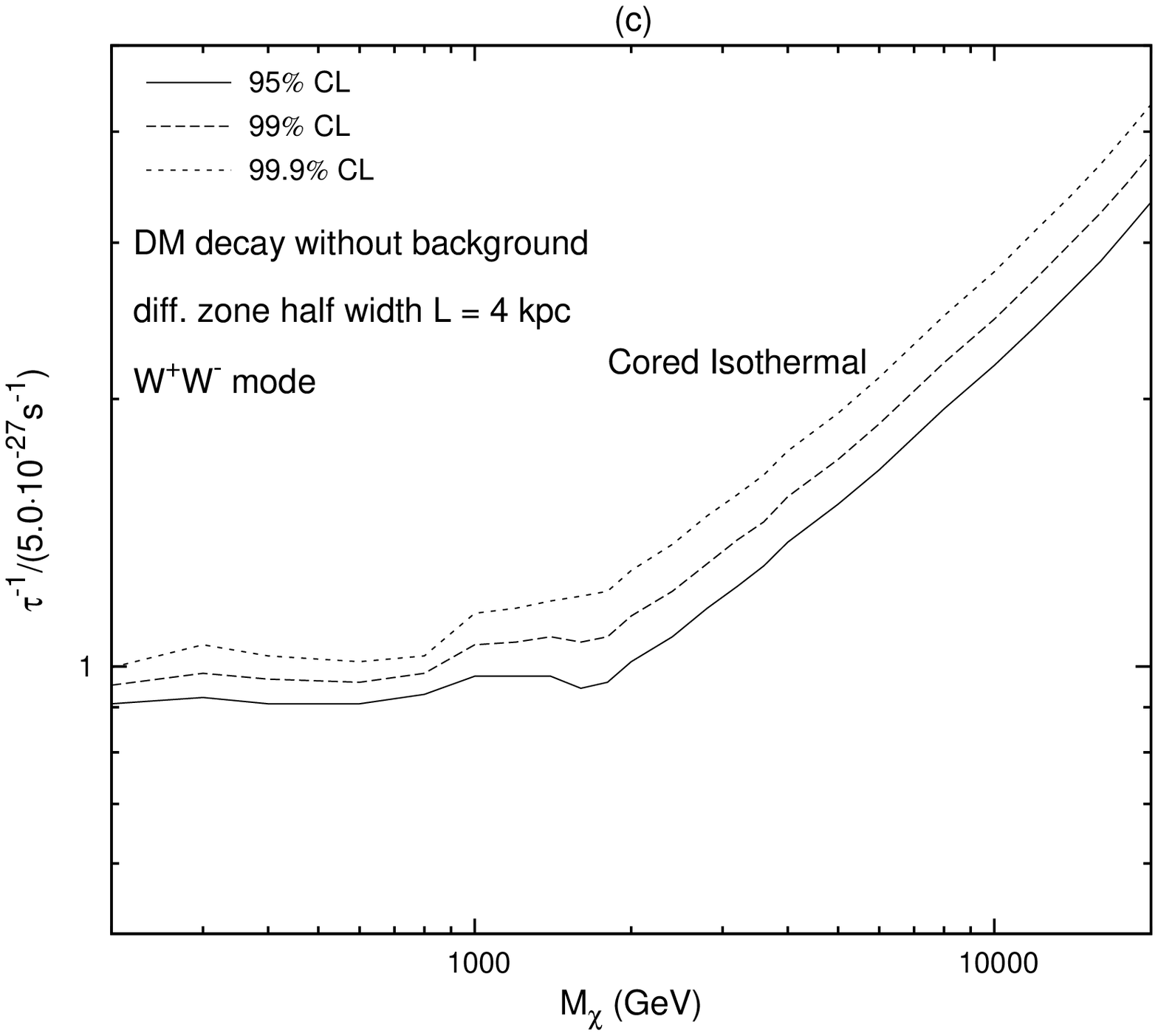}
\hspace{-0.1cm}
\includegraphics[width= 3.20in,angle=0]{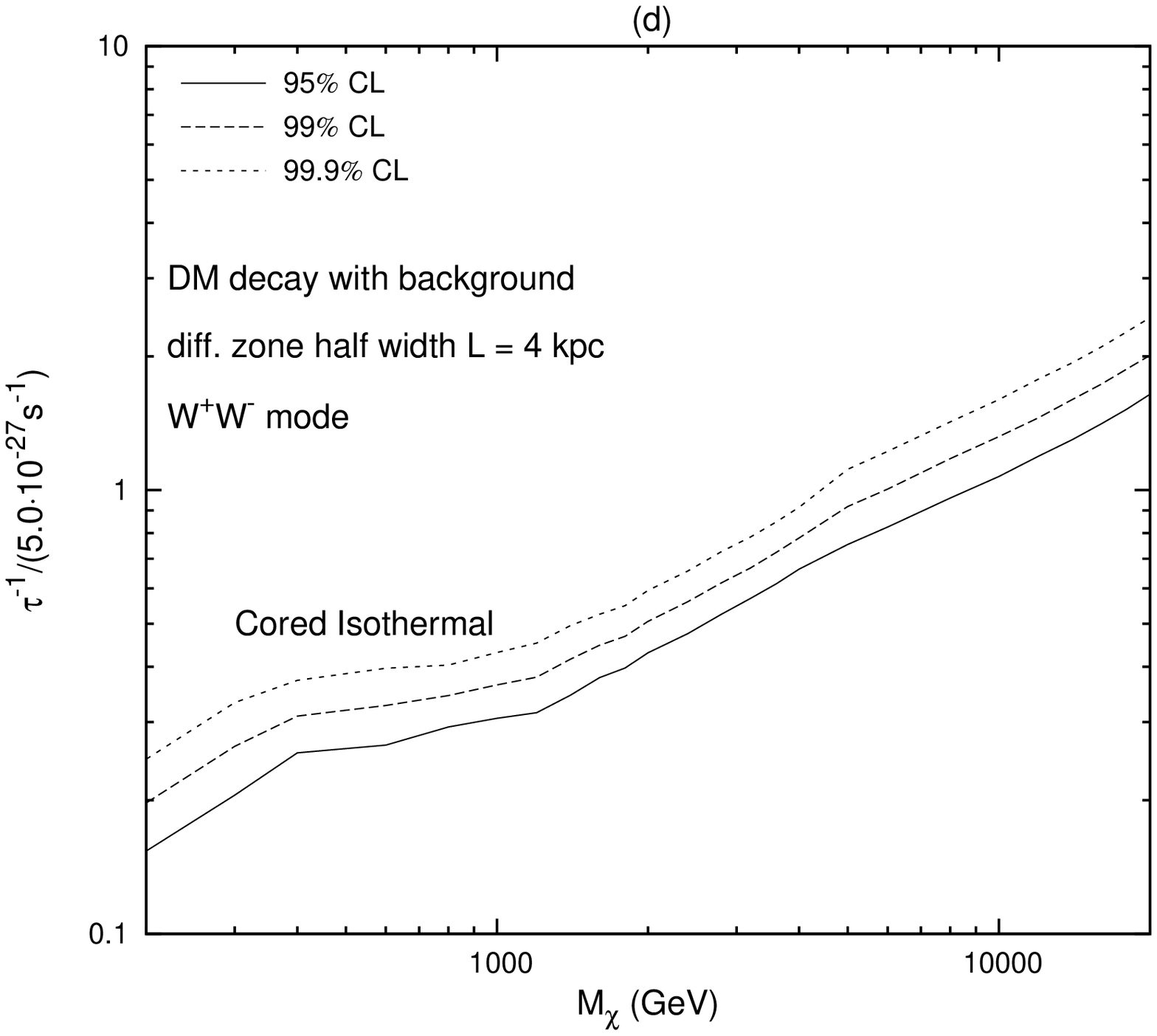}
\caption{As in Fig.~\ref{annihplot} for decaying DM ratios $\tau ^{-1}$/($5.0\times 10^{-27} s^{-1}$).}
\label{decayplot}
\end{figure*}

In figure~\ref{decayplot} we present for the case of decaying DM $\chi \rightarrow W^{+}W^{-}$ the allowed within $95 \%$,  $99 \%$ and $99.9 \%$ C.L. inverse lifetime ratios $\tau ^{-1}$/($5.0\times 10^{-27} s^{-1}$). As with the annihilation scenario we show the results for Einasto profile figure~\ref{decayplot}a,b and for cored isothermal~\ref{decayplot}c,d coming from both the absence of a background~\ref{decayplot}a,c and the presence of it~\ref{decayplot}b,d. The evolution of the upper limits to the mass $m_{\chi}$ is similar as in the case of annihilation (changing $m_{\chi} \rightarrow 2m_{\chi}$). Also as expected the upper limits in the decay lifetimes$^{-1}$ are less sensitive to the DM halo assumptions than in the annihilation case. 

In table~\ref{decay-table} we present the allowed within $95 \%$ C.L. decay inverse lifetime ratios $\tau ^{-1}$/($5.0\times 10^{-27} s^{-1}$) for $m_{\chi} = $ 200 GeV, 2.0 TeV and 20TeV. As in table~\ref{annih-table} we show results for Einasto and cored isothermal profiles, for  $\chi \rightarrow Z Z$, $\chi \rightarrow W^{+} W^{-}$, $\chi \rightarrow b \bar{b}$, with (without) background $\bar{p}$s. 

In table~\ref{diffusion-table} we show the effect of assuming a different diffusion zone thickness ($2L$) in using GALPROP to propagate $\bar{p}$ in the Galaxy. In each case we used a diffusion parametrization that gives local CR nuclei, B/C and sub Fe/Fe fluxes in good agreement with data (see \cite{Simet:2009ne}). As wider diffusion zones include propagation of $\bar{p}$ from a bigger volume of the DM halo, they set tighter constraints on annihilation cross-sections and decay inverse lifetimes. Other anti-proton backgrounds have also been proposed \cite{Donato:2008jk, DiBernardo:2009ku, Bergstrom:1999jc, Bringmann:2006im, Evoli:2008dv, Maurin:2006hy, Maurin:2006ps, Blasi:2009bd} providing upper limits on DM annihilation/decay rates that can be higher by $\sim 2$ (see \cite{Donato:2008jk}), or significantly lower as for the case of \cite{Blasi:2009bd}, where the authors, have studied the significance of the late stages of the evolution of supernova remnants in producing $\bar{p}$ (see also \cite{Blasi:2009hv} for the description of their model). 

In setting the constraints, we ignored DM substructure that enhances the $\bar{p}$ production rate. Including DM substructure results in tighter limits on the annihilation cross-section or decay lifetime. Since $p$ and $\bar{p}$ travel from greater distances compared to $e^{\pm}$ and the boost from substructure in the annihilation case is $\sim 1$ in the inner 10kpc but can grow up to $\sim$$10^{3}$ at $\sim$100kpc from the galactic center\cite{Kamionkowski:2010mi} the exact factor by which our results can vary by including substructure is a more complex question which also depends on the general propagation properties assumed at distances $\sim$100kpc from the galactic center. In the case where the dark disk halo component (seen in recent cosmological simulations including baryons \cite{Read:2008fh, Pato:2010yq}) is significant, the allowed annihilation cross-section upper limits can be increased up to a factor of 6 \cite{Cholis:2010px}.

Also even though $\rho_{0} = 0.4 \; GeV \, cm^{-3}$ that we have used for the local DM density is a value in good agreement with all estimates, current literature\cite{Weber:2009pt, Salucci:2010qr} suggests that it can still be a factor of $\sim 2$ lower or higher. That induces an uncertainty of a factor of $\sim 4$ ($\sim 2$) to the limits that we present for annihilation(decay).

\begin{figure*}[]
\centering\leavevmode
\includegraphics[width=3.20in,angle=0]{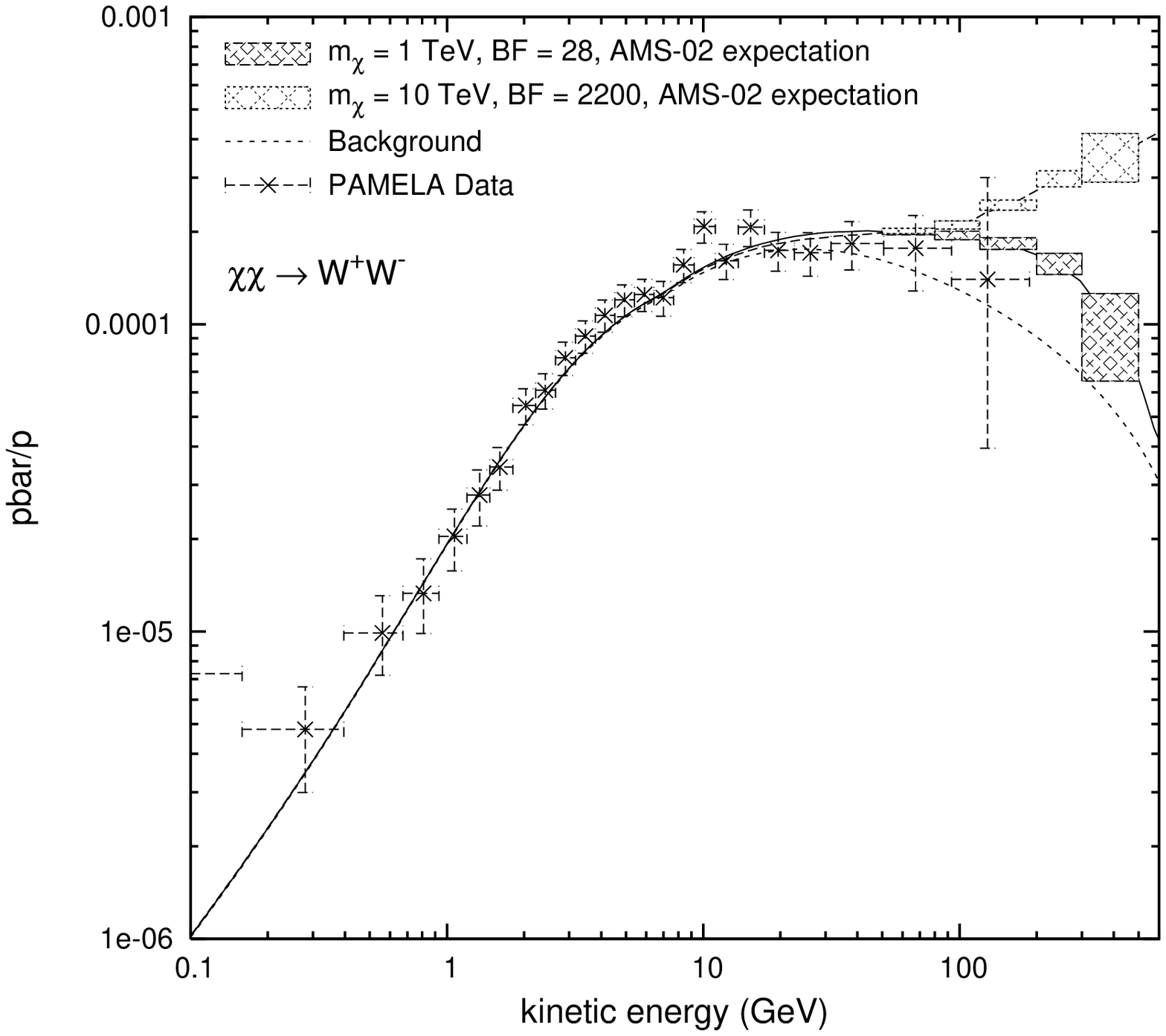}\\
\caption{Fit of $\bar{p}/p$ versus kinetic energy data from  $\chi \chi \rightarrow W^{+}W^{-}$ with  $m_{\chi} = 1.0$ TeV and $m_{\chi} = 10$ TeV. The projected \textit{AMS-02} uncertainties are also shown.}
\label{AMS02plot}
\end{figure*}

In figure~\ref{AMS02plot} we show the $\bar{p}/p$ flux ratio data of \cite{Adriani:2010rc} fitted by our background,\footnote{The $\bar{p}/p$ flux ratio is affected at $E<2$GeV by solar modulation\cite{2002cosp...34E.606L, 2004AdSpR..34..132P} that is taken into account by assuming a modulation potential of 1.1 GeV.} described in~\ref{sec:assumptions} and annihilating DM, $\chi \chi \rightarrow W^{+}W^{-}$ with boost factors BF = 28 for $m_{\chi} = 1.0$ TeV and BF = 2200 for $m_{\chi} = 10$ TeV, where BF = $\langle \sigma \mid v \mid \rangle$/($3.0\times 10^{-26} cm^{3} s^{-1}$). We used an Einasto profile as given in eq.~\ref{eq:Einasto} and our reference propagation assumptions. We also show for the two DM cases presented, the projected \textit{AMS-02} \cite{Casadei:2004kt, Pereira:2007ft} uncertainties on the $\bar{p}/p$ flux ratio after 3 years of data, where the $\bar{p}$ geometrical acceptance is taken to be $A_{\bar{p}}=0.25 m^2 sr$ up to 300 GeV and dropping to 0.04 at 500 GeV \cite{Pato:2010ih, Kounine:2010js, KouninePrivate}. We note also that systematic uncertainties such as a deteriorating energy resolution with increasing energy and at the highest $\bar{p}$ energies proton spill-over from soft scatterings inside the detector can also increase the $\bar{p}$ and $\bar{p}/p$ ratio, placing a limit on the highest energies the $\bar{p}$ measurement can be considered accurate.  
With these uncertainties taken properly into account the \textit{AMS-02} that since May 2011 is on board of the International Space Station, with its large acceptance can provide valuable information to discriminate among various DM scenarios or further constrain them, and  together with its measurements of heavier CR nuclei, decrease the uncertainties in galactic propagation\cite{Pato:2010ih}.

\begin{table}[t]
\begin{tabular}{|cc|cc|cc|cc|}
\hline
Mode& Profile& \multicolumn{2}{|c|}{$m_{\chi} = 100$ GeV} & \multicolumn{2}{|c|}{$m_{\chi} = 1.0$ TeV} &\multicolumn{2}{|c|}{$m_{\chi} = 10$ TeV} \\
\hline \hline
$ZZ$ & Ein. & 1.00 & (16) & 90 & (230) & 5800 & (8300) \\
$W^{+}W^{-}$ & Ein. & 1.03 & (15) & 68 & (160)& 2500 & (5100) \\
$b \bar{b}$ & Ein. & 1.4 & (14) & 81 & (170) & 3800 & (7800) \\
\hline
$ZZ$ & C.I. & 3.2 & (23) & 130 & (340) & 8500 & (12000) \\
$W^{+}W^{-}$ & C.I. & 3.6 & (21) & 97 & (230) & 3600 & (7400) \\
$b \bar{b}$ & C.I. & 3.0 & (22) & 120 & (240) & 5600 & (11000) \\
\hline \hline
\end{tabular}
\caption{Allowed upper limits of ratio $\langle \sigma \mid v \mid \rangle$/($3.0\times 10^{-26} cm^{3} s^{-1}$) within $95 \%$ C.L. for annihilation modes  $\chi \chi \rightarrow Z Z$, $\chi \chi \rightarrow W^{+} W^{-}$, $\chi \chi \rightarrow b \bar{b}$ for Einasto (Ein.) and cored isothermal (C.I.) profiles. Values in parenthesis are calculated ignoring background $\bar{p}$ contribution.}
\label{annih-table}
\end{table}

\begin{table}[t]
\begin{tabular}{|cc|cc|cc|cc|}
\hline
Mode& Profile& \multicolumn{2}{|c|}{$m_{\chi} = 200$ GeV} & \multicolumn{2}{|c|}{$m_{\chi} = 2.0$ TeV} &\multicolumn{2}{|c|}{$m_{\chi} = 20$ TeV} \\
\hline \hline
$ZZ$ & Ein. & 0.11 & (0.71) & 0.43 & (1.01) & 1.6 & (3.5) \\
$W^{+}W^{-}$ & Ein. & 0.15 & (0.66) & 0.42 & (0.99) & 1.6 & (3.2) \\
$b \bar{b}$ & Ein. & 0.15 & (0.67) & 0.42 & (1.04) & 2.5 & (5.0) \\
\hline
$ZZ$ & C.I. & 0.12 & (0.74)& 0.45 & (1.06) & 1.7 & (3.6) \\
$W^{+}W^{-}$ & C.I. & 0.15 & (0.68) & 0.43 & (1.01) & 1.6 & (3.3) \\
$b \bar{b}$ & C.I. & 0.15 & (0.69) & 0.43 & (1.07) & 2.5 & (5.1) \\
\hline \hline
\end{tabular}
\caption{Allowed upper limits of ratio $\tau ^{-1}$/($5.0\times 10^{-27} s^{-1}$) within $95 \%$ C.L. for decay modes  $\chi \rightarrow Z Z$, $\chi \rightarrow W^{+} W^{-}$, $\chi \rightarrow b \bar{b}$ similar to table~\ref{annih-table}.}
\label{decay-table}
\end{table}

\vskip 0.05in

\begin{table}[t]
\begin{tabular}{|cc|cc|cc|cc|}
\hline
Mode& Mass& \multicolumn{2}{|c|}{$L = 2$ kpc} & \multicolumn{2}{|c|}{$L = 4$ kpc} &\multicolumn{2}{|c|}{$L = 10$ kpc} \\
\hline \hline
$W^{+}W^{-}$ annih. & 0.2 & 24 & (84) & 6.6 & (27) & 3.1 & (11) \\
$W^{+}W^{-}$ annih. & 1.0 & 260 & (470) & 68 & (160) & 27 & (63) \\
$W^{+}W^{-}$ annih. & 10 & 7600 & (15000) & 2500 & (5100) & 1000 & (2100) \\
\hline
$W^{+}W^{-}$ decay & 0.4 & 0.44 & (1.5)& 0.25 & (0.84) & 0.11 & (0.38) \\
$W^{+}W^{-}$ decay & 2.0 & 0.72 & (2.9) & 0.42 & (1.03) & 0.19 & (0.56) \\
$W^{+}W^{-}$ decay & 20 & 2.8 & (7.9) & 1.6 & (4.4) & 0.69 & (1.5) \\
\hline \hline
\end{tabular}
\caption{Allowed within $95 \%$ C.L. upper limits of annihilation to $W^{+}W^{-}$ cross-section ratio $\langle \sigma \mid v \mid \rangle$/($3.0\times 10^{-26} cm^{3} s^{-1}$) and decay to $W^{+}W^{-}$ ratio $\tau ^{-1}$/($5.0\times 10^{-27} s^{-1}$) for varying mass (TeV) and diffusion zone half width $L$.}
\label{diffusion-table}
\end{table}

%%%%%%%%%%%%%%%%%%%%%%%%%%%%%%%%%%%%%%%%%%%%%%%%%%%%%%%%%%%%%%%%%%%%%%%%%%
%%%%%%%%%%%%%%%%%%%%%%%%%%%%%%%%%%%%%%%%%%%%%%%%%%%%%%%%%%%%%%%%%%%%%%%%%%
 
\section{Conclusions}
\label{sec:conclusions}

Using the new $\bar{p}$ \textit{PAMELA} data \cite{Adriani:2010rc} 
we have provided a set of constraints on DM annihilation cross-section, decay lifetimes, for cases where the DM annihilates/decays to $ZZ$, $W^{+}W^{-}$, $b \bar{b}$ producing significant fluxes of high energy $\bar{p}$. The new constraints are in agreement with analysis of the older \textit{PAMELA} data as in \cite{Donato:2008jk}. Uncertainties in galactic propagation and the shape and local normalization of the DM halo density allow the constraints in the annihilation cross-section/decay lifetimes to vary typically by factors of 2-5 but is some cases up to $\sim 10$. For DM models that can explain \textit{PAMELA} positron fraction \cite{Adriani:2008zr, Adriani:2010ib} from either a boosted annihilation rate or an appropriate decay rate, the new $\bar{p}$ data still suggest in general DM annihilating/decaying preferentially to leptons than hadrons\footnote{We clarify that both the absence of a hardening of the antiproton flux and the rising positron fraction(together with a hardening of the $e^{+}+e^{-}$ flux) can be explained by more ``conventional'' sources \cite{Hooper:2008kg, Profumo:2008ms, Malyshev:2009tw}.}. Based on our most conservative upper limits analysis a case of 200 GeV LSP wino as suggested by \cite{Kane:2009if} is already tightly constrained by the new data. Yet wino masses slightly larger as still allowed, thus models where DM can annihilate via hadronic modes \cite{Grajek:2008pg} are still not excluded. Including background antiprotons though increases the constrained parameter space to below 300 GeV, at which point models such as that of \cite{Kane:2009if} can not explain the rise of positron fraction; unless other sources such as pulsars are included (for a recent analysis of constraints on DM annihilating models with a wide range of assumed propagation conditions see \cite{Evoli:2011id}).  
Diffuse gamma-rays from the galactic center and galactic ridge \cite{Aharonian:2006wh, Aharonian:2006au} and from inner parts of the Galaxy measured by \textit{Fermi}-LAT have been shown to set constraints on DM annihilating via hadronic modes (see for instance \cite{Meade:2009rb, Cirelli:2009dv, Crocker:2010gy}). Finally the \textit{AMS-02} on board the ISS, due to its high acceptance will be able to measure the $\bar{p}/p$ and $\bar{p}$ spectrum up to at least $\sim 300$ GeV allowing to further constrain or detect DM annihilating/decaying via modes that produce significant numbers of hadrons.

\vskip 0.2 in
\noindent {\bf Acknowledgments}
IC has been supported by DOE OJI grant \# DE-FG02-06ER41417, and also by the Mark Leslie Graduate Assistantship.  The author would like to thank Mirko Boezio, Lisa Goodenough and Neal Weiner for valuable discussions and insight they have provided.
\vskip 0.05in
\noindent

%%%%%%%%%%%

\bibliography{antiprotonsletter}
\bibliographystyle{apsrev}

\end{document}